\documentclass[useAMS,usenatbib,aas_macros]{mn2e}
\usepackage{aas_macros}

\usepackage{color}

\newcommand{\equ}[1]{eq.~(\ref{eq:#1})}

\newcommand{\Equ}[1]{Eq.~(\ref{eq:#1})}

\newcommand{\equnp}[1]{eq.~\ref{eq:#1}}
\newcommand{\se}[1]{\S\ref{sec:#1}}

\newcommand{\tab}[1]{Table~\ref{tab:#1}}
\newcommand{\Tab}[1]{Table~\ref{tab:#1}}
\newcommand{\be}{\begin{equation}}
\newcommand{\ee}{\end{equation}}
\newcommand{\bea}{\begin{eqnarray}}
\newcommand{\eea}{\end{eqnarray}}
\def\ra{\rangle}
\def\la{\langle}

\def\bul{\smallskip\noindent$\bullet$\ }

\newcommand{\msun}{M_\odot}

\newcommand{\lsun}{L_\odot}
\newcommand{\ifm}[1]{\relax\ifmmode#1\else$\mathsurround=0pt #1$\fi}
\newcommand{\kms}{\ifmmode\,{\rm km}\,{\rm s}^{-1}\else km$\,$s$^{-1}$\fi}

\newcommand{\kpc}{\,{\rm kpc}}
\newcommand{\pc}{\,{\rm pc}}
\newcommand{\Gyr}{\,{\rm Gyr}}

\newcommand{\Myr}{\,{\rm Myr}}

\newcommand{\gcms}{\,{\rm g\,cm}^{-2}}

\newcommand{\ergs}{\,{\rm erg\,s}^{-1}}

\newcommand{\ltsima}{$\; \buildrel < \over \sim \;$}
\newcommand{\lsim}{\lower.5ex\hbox{\ltsima}}
\newcommand{\gtsima}{$\; \buildrel > \over \sim \;$}
\newcommand{\gsim}{\lower.5ex\hbox{\gtsima}}

\def\sy{\,M_\odot\, {\rm yr}^{-1}}

\def\Rv{{R_{\rm v}}}
\def\Vv{{V_{\rm v}}}

\def\Mg{{M_{\rm g}}}
\def\Ms{{M_\star}}
\def\Mdots{{\dot{M}_\star}}

\def\Md{M_{\rm d}}
\def\Rd{{R_{\rm d}}}
\def\Rt{{R_{\rm T}}}

\def\td{{t_{\rm d}}}
\def\Vd{{V_{\rm d}}}

\def\sigd{\sigma_{\rm d}}

\def\fg{{f_{\rm g}}}

\def\Mc{{M_{\rm c}}}
\def\Rc{{R_{\rm c}}}

\def\Vc{V}
\def\Vrot{V_{\rm rot}}

\def\rhod{\rho_{\rm d}}

\def\Mdotw{{\dot{M}_{\rm w}}}

\def\tff{{t_{\rm ff}}}

\def\tdep{{t_{\rm dep}}}
\def\tmig{{t_{\rm mig}}}

\def\epsf{{\epsilon_{\rm ff}}}
\def\eps2{{\epsilon_{-2}}}
\def\epss{{\epsilon_{\star}}}

\def\ftrap{f_{\rm trap}}

\def\sfr{{\dot{M}_{\star}}}
\def\SFR{{\dot{M}_{\star}}}
\def\Mdotw{\dot{M}_{\rm w}}

\def\ftrap{f_{\rm trap}}
\def\pdotin{\dot{p}_{\rm in}}
\def\pdotw{\dot{p}_{\rm w}}

\def\Vc{{V_{\rm c}}}
\def\Vw{{V_{\rm w}}}
\def\Vl{{V_{\rm L}}}
\def\Vin{{V_{\rm in}}}
\def\Vrad{{V_{\rm rad}}}
\def\Vps{{V_{\rm ps}}}
\def\Vsn{{V_{\rm sn}}}
\def\Vms{{V_{\rm ms}}}

\def\fej{{\psi_{\rm ej}}}
\def\fin{{\psi_{\rm in}}}
\def\fw{{\psi_{\rm w}}}
\def\tsfr{t_{\rm sfr}}
\def\tw{t_{\rm w}}
\def\tac{t_{\rm ac}}

\def\n0{n_0}
\def\n1{n_1}

\newcommand{\Rexp}{R_{\rm exp}}

\title[Outflows in Giant Clumps]
{Steady Outflows in Giant Clumps of High-z Disk Galaxies During Migration and
Growth by Accretion}

\author[A. Dekel and M.R. Krumholz]
{Avishai Dekel$^{1}$\thanks{avishai.dekel@mail.huji.ac.il}
 and Mark R. Krumholz$^{2}$\thanks{krumholz@ucolick.org}
\\ \\
$^1$Racah Institute of Physics, The Hebrew University, Jerusalem 91904 Israel\\
$^2$Department of Astronomy \& Astrophysics, University of California, Santa
Cruz, CA 95060, USA}

\begin{document}

\large

\pagerange{\pageref{firstpage}--\pageref{lastpage}} \pubyear{2013}

\maketitle

\label{firstpage}

\begin{abstract}
We predict the evolution of giant clumps undergoing star-driven outflows in 
high-$z$ gravitationally unstable disc galaxies. We find that the mass loss is 
expected to occur through a steady wind over many tens of free-fall times 
($\tff \sim 10\Myr$) rather than by an explosive disruption in one or a few 
$\tff$. Our analysis is based on the finding from simulations that radiation 
trapping is negligible because it destabilizes the wind \citep{krumholz12c, 
krumholz13}. Each photon can therefore contribute to the wind momentum only 
once, so the radiative force is limited to $L/c$. When combining radiation, 
protostellar and main-sequence winds, and supernovae, we estimate the total 
direct injection rate of momentum into the outflow to be $2.5\,L/c$. The 
adiabatic phase of supernovae and main-sequence winds can double this rate. The
resulting outflow mass-loading factor is of order unity, and if the clumps were
to deplete their gas the timescale would have been a few disc orbital times, to
end with half the original clump mass in stars. However, the clump migration 
time to the disc centre is on the order of an orbital time, about $250 \Myr$, 
so the clumps are expected to complete their migration prior to depletion. 
Furthermore, the clumps are expected to double their mass in a disc orbital 
time by accretion from the disc and clump-clump mergers, so their mass actually
grows in time and with decreasing radius. From the 6-7 giant clumps with 
observed outflows, 5 are consistent with these predictions, and one has a much
higher mass-loading factor and momentum injection rate. The latter either 
indicates that the estimated outflow is an overestimate (within the 1-$\sigma$ 
error), that the SFR has dropped since the time when the outflow was launched, 
or that the driving mechanism is different, e.g.~supernova feedback in a cavity
generated by the other feedbacks.
\end{abstract}

\begin{keywords}
{
%dark matter ---
%galaxies: elliptical ---
%galaxies: evolution ---
galaxies: formation ---
%galaxies: haloes ---
galaxies: ISM ---
%galaxies: mergers
galaxies: spiral ---
galaxies: star clusters ---
ISM: jets and outflows ---
stars: formation
}
\end{keywords}

%%%%%%%%%%%%%%%%%%%% 1
\section{Introduction}
\label{sec:intro}

In our developing picture of violent disc instability (VDI) at high redshift,
the gas-rich discs fed by cosmological streams give birth to giant baryonic 
clumps that are the sites for intense star formation. 
The clumps are expected to migrate toward the disc centre on an orbital time 
scale where they coalesce into the central bulge
\citep{noguchi99,immeli04_b,bournaud07c,elmegreen08a,
genzel08,dsc09,agertz09,cdb10}.
This was proposed as a mechanism for the formation of galactic spheroids,
in parallel with the traditional scenario of spheroid formation by mergers
\citep{genzel08,dsc09}
as well as a scenario for the formation of globular clusters
\citep{shapiro10}, and for feeding the central black holes 
\citep{bournaud11,bournaud12}.
However, stellar feedback can generate outflows from the clumps 
\citep{murray10,kd10}.
These outflows were assumed to be very intense on a timescale of a few 
free-fall times
and thus lead to significant mass loss and possibly to clump disruption
\citep{murray10,hopkins12c,genel12a}. 
\citet{murray10} argued that the high-z giant clumps are likely to be disrupted
by momentum-driven feedback, as are their smaller counterpart molecular clouds
in the Milky Way at low redshift, 
but \citet{kd10} pointed out that this would be possible
only if the efficiency of star-formation per free-fall time $\epsf$
is significantly higher than the value implied by observations of both 
nearby and high-$z$ galaxies \citep{krumholz07e,krumholz12a},
namely the value associated with the Kennicutt-Schmidt (KS) relation.

\smallskip
\citet{genzel11} reported pioneering observational evidence for outflows
from giant clumps in five $z \sim 2$ galaxies. 
The SFR is estimated from the H$_\alpha$ luminosity. 
The clump properties of radius and characteristic velocity are measured
directly,
or alternatively the gas mass is derived from the SFR assuming that the
KS relation observed on sub-galactic scales at $z=0$
\citep{krumholz07e, krumholz12a}
and on galactic scales at $z\sim 2$
\citep{genzel10b,daddi10_sfr,tacconi10,daddi10_co,tacconi13}
also holds within individual giant clumps at $z\sim 2$ 
\citep[see preliminary results by][]{freundlich13}.
%\adb{
The outflow velocity is evaluated based on both the centroid blue-shift and the
width of the broad-line component of the H$_\alpha$ emission, and the mass
outflow rate is estimated using a variety of alternative models.
%}
% and mass outflow rate are evaluated from
%the blue-shift  and t relative intensity of the broad component of 
%the H$_\alpha$ emission using a variety of alternative models.
Based on these observations, \citet{genzel11} estimated that
the typical clumps in their sample drive winds of mass loading factor
$\eta \sim 1$, namely a mass outflow rate that is comparable to the SFR.
These winds are expected to deplete the clump gas in 
several hundred $\Myr$ after turning about half the clump mass into stars.
Two extreme cases, both in the same galaxy, ZC406690, 
indicate stronger outflows,
with $\eta \sim 3-7$ and depletion times of $(100-200)\Myr$ with
only 12-25\% of the original clump mass turning into stars.
In the typical clumps, the actual driving force of the outflows is
$(2-4)\,L/c$, where $L/c$ is the contribution of a single-scatter radiation
force, while in the most extreme outflow it is estimated to be as large
as $34\,L/c$ (though with a very large uncertainty). 

\smallskip
\citet{newman12} performed follow-up observations at higher resolution on
the galaxy with the extreme clumps, ZC406690. They compared
two clumps in this galaxy, both of which are driving winds, but with very 
different
properties. One of the clumps shows a considerably larger mass, energy, and
momentum flux than the other. They propose that these two clumps represent
different evolutionary stages of the same phenomenon, and that the more
energetic of the two outflows cannot easily be explained by any of the wind
launching mechanisms that have been proposed in the literature.

\smallskip
In this paper we seek to provide a unified framework for comparing different
potential outflow launching mechanisms, and then use this framework to
predict the outflow properties expected from stellar feedback, and
understand what can be learned from the observations conducted to date.
We consider the momentum injected into the wind by momentum-conserving stellar
feedback mechanisms, and by the more energy-conserving supernova feedback.
We predict the expected mass loading factor and momentum injection efficiency.
We focus in particular on the question of whether the migrating clumps arrive
at the centre massive and intact or lose most of their mass to outflows 
while still in the disc. 

\smallskip
In \se{frame} we develop a simple theoretical framework for dealing with 
the momentum that drives outflows from star-forming clumps.
We present in comparison the timescale for clump migration and the 
accretion rate into the clumps during migration.
In \se{budget} we go through the momentum budget for the outflows.
In \se{implications} we address the implications for the evolution
of high-$z$ giant clumps.
In \se{obs} we compare the observational estimates to the predictions,
and discuss the implications on the outflow driving mechanisms.
In \se{conc} we conclude our results and discuss them.

%%%%%%%%%%%%%%%%%%%
\section{Theoretical Framework}
\label{sec:frame}

\subsection{Momentum versus Energy Feedback}

Newly formed stars generate outflows by injecting momentum and energy
into the interstellar gas.  Our goal in this section is to develop a 
basic theoretical machinery to describe this phenomenon.
We first address the roles of momentum and energy in this context,
and clarify the terminology of momentum-driven versus energy-driven 
feedback.

\smallskip
The central conceptual challenge is that cool interstellar gas
is highly dissipative, so energy is always lost to radiative processes. Indeed,
in cold gas the cooling time is almost always short compared to dynamical
timescales. Thus in launching a wind, what we are in the end always concerned
with is the amount of momentum that is transferred by stellar feedback to the
gas.\footnote{Note that when considering a spherical outflowing shell 
one refers to the overall momentum in the radial direction, which
is not necessarily conserved.}
We can address two extreme ways for this transfer to occur. 
First, \textit{momentum-conserving} (PC) transfer, where ejecta from stars
(photons, winds, supernova ejecta) collide with the ISM inelastically,
transferring its momentum but losing some of its energy. 
Second, \textit{energy-conserving} (EC) transfer, where stars heat the
interstellar material, either radiatively or via shocks, to temperatures high
enough that the cooling time becomes much longer than the dynamical time. When
this happens the hot gas expands adiabatically, transferring momentum to the
cool phases of the ISM as it does so.
We generally refer to the former
mechanism for launching a wind as momentum-driven and the latter as
energy-driven, but this is a somewhat misleading nomenclature, because the
rapid cooling in the cold phases of the ISM implies that in either case what
ultimately matters is the momentum transferred to the cold gas.

\smallskip
The energy-conserving case is in general much more efficient. 
To see this, consider a source
of ejecta with outflow velocity $V_{\rm s}$ ($V_{\rm s}=c$ for radiation)
and mass flow rate $\dot{M}_{\rm s}$ (or the equivalent energy outflow rate
for radiation).
In the PC case, after time $t$, the ejecta has pushed
a wind of mass $M_{\rm p}$ and velocity $V_{\rm p}$ obeying
\be
M_{\rm p} V_{\rm p} = \dot{M}_{\rm s} t V_{\rm s} \, .
\ee
In the EC case, the ejecta has pushed
a wind of mass $M_{\rm e}$ and velocity $V_{\rm e}$ obeying
\be
M_{\rm e} V_{\rm e}^2 = \dot{M}_{\rm s} t V_{\rm s}^2 \, .
\ee
The ratio of wind energies between the EC and PC cases is
\be
\frac{E_{\rm e}}{E_{\rm p}} 
= \frac{M_{\rm e} V_{\rm e}^2}{M_{\rm p} V_{\rm p}^2}
= \frac {V_{\rm s}}{V_{\rm p}}
= \frac{M_{\rm p}}{\dot{M}_{\rm s} t}
\gg 1 \, ,
\ee
and, more importantly, the corresponding ratio of momenta is
\be
\frac{P_{\rm e}}{P_{\rm p}}
= \frac{M_{\rm e} V_{\rm e}}{M_{\rm p} V_{\rm p}}
= \frac {V_{\rm s}}{V_{\rm e}}
= \left( \frac{M_{\rm e}}{\dot{M}_{\rm s} t} \right)^{1/2}
\gg 1 \, .
\label{eq:p_ratio}
\ee
Both ratios are much larger than unity as long as the wind mass
is much larger than the mass of the direct ejecta from the source,
and the wind velocity is much smaller than the original ejecta velocity.
%\adb{The latter is obvious when the source is photons moving at the speed 
%of light.}
From \equ{p_ratio} we learn that the efficiency of injecting momentum 
into the wind, $\psi = M V/\dot{M}_{\rm s} t V_{\rm s}$,
is much larger in the EC case than in the PC case.
It turns out that radiative-pressure and stellar winds are PC mechanisms,
while the expanding supernova ejecta have an early adiabatic phase that 
makes them closer to EC.

\smallskip
%\mkb{
It is worth pausing for a moment to elaborate on how it is possible for
the radial momentum imparted to the wind to greatly exceed that provided by
the source when the flow is energy-conserving, and what distinguishes the
EC and PC cases. The characteristic signature of the EC case is the presence
of some mechanism that carries information 
and forces between different parts of the expanding shell of swept-up gas, allowing
them to push off one another and thereby greatly increase their radial momenta
while leaving the vector momentum of the shell as zero. The mechanism
responsible may be sound waves traveling through hot gas that communicate
forces via gas pressure, it may be photons bouncing from one
side of a spherical shell to the other that carry information via radiation pressure,
or it may be something else, as long as
that something allows forces to be transmitted from one side of the expanding
shell to the other. In contrast, the distinguishing feature of PC flows is that
there is no causal communication between different parts of the expanding
shell, and as a result they cannot push off each other and increase their
radial momenta.
%}

%------------------------
\subsection{Momentum Injected by Star Formation}

Star formation in clumps, either giant clumps at $z\sim 2$ or much smaller
star-forming clumps in the nearby universe, generally does not continue until
it has exhausted the available gas supply into stars. 
Instead, it ends after some fraction $\epss$ of the initial gas has been 
converted to stars, 
\be
\epss \equiv \frac{M_\star}{\Mc} \, .
\ee
%while the rest of the gas is expelled by the winds 
%generated by the newly formed stars.
%\mkb{
The remaining gas is either directly expelled by the winds, or is
removed by tidal stripping or simply drifts off after the winds have eroded
the clump mass enough to render it unbound.
%}
The quantity $\epss$ is commonly referred to as the
\textit{star formation efficiency}. 
The time of final \textit{gas depletion} is $\tdep = M_\star/\dot{M}_\star$,
where $\dot{M}_\star$ is the star formation rate (SFR) 
and $M_\star$ is the final stellar mass at $\tdep$.
This expulsion can occur via a gradual wind that removes mass from the clump
continuously as it forms stars, via an explosive event that removes the bulk of
the mass on a timescale comparable to or smaller than the clump dynamical time,
or some combination of the two. Our goal in this section is to develop some
basic theoretical machinery to describe this phenomenon. 

\smallskip
Let $\Mc$ and $\Rc$ be the mass and radius of a star-forming clump, and let
\be
\tff=\sqrt{\frac{\Rc^3}{G \Mc}} = \frac{G\Mc}{\Vc^3} \, ,
\quad
\Vc = \sqrt{\frac{G \Mc}{\Rc}} \, .
\label{eq:tff}
\ee
be the corresponding  
free-fall time and characteristic velocity, respectively, 
where we are dropping factors of order unity for simplicity.\footnote{The
expression $\tff = \sqrt{3\pi/(32 G \rho)}$, where $\rho$ is the mean density
within the clump, is $\sqrt{8}/\pi$ times the expression in \equ{tff}.} 
The crossing time $\Rc/\Vc$ is of the same order as $\tff$, 
and the escape speed is of the same order as $\Vc$.\footnote{For an
object with zero pressure and magnetic field the escape speed is $\sqrt{2}\Vc$,
but observed molecular clouds in the local universe are only magnetically
supercritical by factors of $\sim 2$ \citep[e.g.][]{troland08a}, corresponding
to a reduction in the escape speed to $\Vc$. There are no direct measurements
of magnetic field strengths in high redshift giant clumps, but numerical 
simulations
of magnetized turbulence suggest that a turbulent dynamo can rapidly amplify
an initially sub-Alfv\'{e}nic field to an Alfv\'{e}n Mach number of order unity \citep{stone98a}.
If this occurs in giant clumps at high $z$, the escape speed should be 
reduced similarly for them.}
The instantaneous SFR is written as 
\be
\SFR = \epsf \frac{\Mc }{\tff} = \epsf G^{-1} \Vc^3 \, .
\label{eq:sfr}
\ee
The corresponding SFR timescale is
\be
\tsfr \equiv \frac{\Mc}{\sfr} \simeq \epsf^{-1}\tff \, .
\label{eq:tsfr}
\ee
The quantity $\epsf$ is known as the \textit{SFR efficiency} 
per free-fall time, or the rate efficiency, to distinguish it from the overall 
star formation efficiency $\epss$. 
Observations of star formation in a wide variety of environments at a
wide variety of redshifts strongly constrain that $\epsf \sim 0.01$
\citep{krumholz07e, krumholz12a}. 
In particular, there are indications from observed CO that the same relation
with the same $\epsf$ is also valid at $z \sim 2$
\citep{daddi10_sfr,tacconi13}.
In our toy model we assume that the SFR is roughly constant throughout
the clump lifetime. 
%\adb{
This is consistent with cosmological simulations,
where the SFR in clumps does not show a systematic variation with distance 
from the disc centre despite the continuous clump migration inward
\citep{mandelker13}.
%}
We have replaced in \equ{sfr} the instantaneous gas mass $\Mg$ by  
the total initial clump mass $\Mc$. This should be a close overestimate 
as long as the clump is still far from its gas depletion time.
When estimating the depletion time, we will
% assume $\Mg \simeq 0.5\Mc$ (see below). 
%\adb{
replace $\Mc$ in \equ{sfr} by $0.5\,\Mc$, 
which will make $\tsfr$ larger by a factor of 2.
%} 

\smallskip
The stars that form inject momentum into the remaining gas at a rate 
$\pdotin$, in the radial direction.
In this section we will not distinguish between the energy-driven and
momentum-driven routes for producing this momentum. Since feedback is generally
dominated by massive stars, one can appeal to the ``old stars" limit 
\citep{kd10}, where we are concerned with timescales longer than the 
$\sim 4$ Myr lifetime of a massive star\footnote{The 
old stars limit almost certainly applies to giant clumps at $z\sim 2$, since
these have $\tff \gg 4$ Myr \citep{kd10}. It probably applies to giant
molecular clouds in the local universe as well, since, although these have
$\tff \sim 4$ Myr, the best observational estimates of GMC lifetimes are $\sim
30$ Myr \citep{fukui09a}; this is significantly uncertain, however. The old
stars limit does not apply to smaller-scale structures seen in our galaxy (see
\citealt{krumholz12a} for a more thorough discussion).}.
In this limit, the number of massive stars generating
feedback at a given time is simply proportional to the SFR,
so we can write 
\be
\pdotin  \equiv \Vin\SFR \, , \quad
\Vin \equiv \fin \Vl \, .
\label{eq:Vin}
\ee
The velocity $\Vl$ characterizes the momentum carried by the stellar
radiation field,
\be
\frac{L}{c} = \Vl \SFR \, ,
\label{eq:L}
\ee
where $L$ is the luminosity produced by these stars.
As discussed in Section \ref{sec:stellarrad}, this quantity is
\be
\Vl = 190\kms \, ,
\label{eq:Vl}
\ee
corresponding to an energy production of 
$c \Vl = 5.7\times 10^{17}$ erg g$^{-1}$.
The dimensionless parameter $\fin$ is the {\it momentum injection factor}; 
it measures the multiplicative factor 
by which the actual injected momentum is higher than that one would obtain
if stellar radiation were the only source of momentum, and if every
photon were absorbed only once before escaping. For convenience, we sometimes 
express it in terms of an effective {\it trapping factor\,} (see below)
\be
\fin \equiv 1+ \ftrap \, .  
\label{eq:ftrap}
\ee
 
%---------------------
\subsection{Wind Properties}
 
If the momentum injected by stars is able to raise material to speeds 
$\Vw \gsim \Vc$, this material may be driven off the clump in a steady wind.
Observed outflows from clumps suggest that $\Vw/\Vc$ is of order a few.
Given the available supply of momentum, \equ{Vin}, the actual momentum of the
wind is
\be
\pdotw = \fej \Vin \SFR \, ,
%\equiv \fw \Vl \SFR \, ,
\label{eq:fej}
\ee
where $\fej$ is the {\it ejection efficiency\,} representing the fraction
of the injected stellar momentum that goes into the wind.
The ejection efficiency could be lower than unity if some material is
raised to speeds below $\Vc$ and thus does not escape.  Similarly, if the
momentum injection is spatially distributed rather than point-like, there may 
be some cancellation of the momenta injected at different positions, again
producing $\fej <1$.
We define for simplicity the overall 
{\it momentum efficiency factor\,} in driving the wind,
\be
\fw \equiv \fej \fin = \frac{\pdotw}{L/c} \, ,
\label{eq:fw}
\ee
as the factor representing the ratio of actual momentum in the wind 
to the momentum carried by the radiation when each photon is counted once.
This quantity can be addressed observationally, as opposed to $\fin$ and
$\fej$.

\smallskip
The wind mass flow rate can be extracted from the wind momentum and velocity
via 
\be
\Mdotw \Vw = \pdotw = \fw \Vl \SFR 
% = G^{-1} \epsf \fw \Vl \Vc^3 
\, .
\label{eq:Mdotw}
\ee
The {\it mass loading factor\,} of the wind is 
\be
\eta \equiv \frac{\Mdotw}{\SFR} = \fw \frac{\Vl}{\Vw} \, .
\label{eq:eta}
\ee
The corresponding timescale for mass loss by outflow is then
\be
\tw \equiv \frac{\Mc}{\Mdotw} \simeq \eta^{-1} \epsf^{-1} \tff \, .
\label{eq:tw}
\ee
For $\eta$ of order unity, this timescale is comparable to the SFR timescale, 
\equ{tsfr}.

\smallskip
Note from $\pdotw = \Mdotw \Vw$ that the mass-loss rate for a given
momentum budget is maximized if the wind {\it velocity factor}
\be
\nu \equiv \frac{\Vw}{\Vc} 
\label{eq:nu}
\ee
is as close to unity (from above) as possible,
i.e.~if the ejected gas is raised to the lowest possible velocity consistent
with escape. 
If $\nu$ is a constant determined by the stellar momentum-driven ejection 
mechanism and is independent of the clump escape velocity (as indicated 
observationally, $\nu \sim 3$, see \se{obs}), 
then $\eta$ is inversely proportional to $\Vc$,
namely the wind mass-loading factor is larger for less massive clumps.

\smallskip
The extreme limit of this phenomenon, as considered by \citet{fall10a} and
\citet{kd10}, is an \textit{explosive ejection}, which
occurs when the momentum or energy injection is sufficient to
raise the entire gas mass to speeds $\gsim \Vc$ in a time of order $\tff$. In
this case the feedback is likely to sweep up all the material in the clump and
eject it explosively, halting any further star formation. 
%\mkb{
The condition for this to occur is that $\pdotw \tff \sim \Mg \Vc$, which,
using \equ{Mdotw}, reduces to the condition that 
\be
\frac{\dot{M}_*}{(\Mg/\tff)}  = \epsf \gsim \frac{1}{\fw}\frac{\Vc}{\Vl}.
\label{eq:explosive}
\ee
As pointed out by \citet{kd10}, this condition is extremely difficult to
satisfy in giant clumps. Observations constrain $\epsf \sim 0.01$, and,
as we will see below, $\Vc / \Vl \sim 0.5$ for giant clumps. Thus
achieving explosive ejection requires either that $\fw \sim 10-100$,
that $\epsf$ in giant clumps exceed the observationally-inferred values
of $\sim 0.01$ by a factor of $\sim 10-100$, or some combination of both.
\subsection{Clump Depletion}

Given our calculated mass loss rates, we can also consider the implications for
the star formation efficiency and lifetime of star forming clumps. First
consider a clump that never experiences explosive ejection, and is simply
eroded by a combination of star formation and a steady wind 
%\adb{
at constant rates.
%}
The star formation efficiency at the time of gas depletion will be 
\be
\epss 
= \frac{\dot{M}_*}{\dot{M}_* + \dot{M}_{\rm w}}
= \frac{1}{1+\eta} \, .
%= \frac{1}{1+ \fej \Vin/\Vw} \, ,
\label{eq:epss}
\ee
%with $\eta$ the mass loading factor from \equ{eta}.
Conversely, a clump that has no steady wind ($\psi_{\rm ej} = 0$)
%\mkb{
and also does not satisfy the criterion for explosive disruption, 
\equ{explosive}, will eventually turn itself completely into stars, or
will undergo explosive disruption at late times when the amount of
gas is reduced to the point where the clump is no longer in the
old stars limit. By this point, however, it will have already turned
the great majority of its mass into stars.
%}
A more realistic
scenario is that a clump experiences a steady wind during its
life %\mkb{
and its 
%lifetime is 
star formation efficiency is given by \equ{epss}.
%}
% will undergo
%explosive disruption and stop forming stars once $\Mg/\Mc$ reaches the critical
%value given in equation (\ref{eq:mgcrit}). The resulting star formation
%efficiency will be $\epss = 1-\Mg/\Mc$, which evaluates to 
%\be
%\label{eq:epss_explosive}
%\epsilon_{*,\rm exp} \simeq 1 - 0.5\, \epsf\,\fw\, \frac{\Vl}{\Vc} \, .
%\label{eq:epssexp}
%\ee
%With $\fw$ of order unity, and $\epsf \sim 0.01$, this is likely to be just a
%bit below unity.

%\smallskip
%A more realistic scenario is that a clump experiences a steady wind during its
%life and then experiences explosive ejection once its mass has been eroded
%sufficiently by the wind and by star formation. If the mass at which explosive
%ejection occurs is small compared to the initial mass, then this will be nearly
%identical to the simple steady wind scenario, while if it is 
%%larger than or
%comparable to the initial mass, the situation will be close to the explosive
%ejection scenario. Given the crudeness of our calculations, a reasonable way of
%interpolating between these regimes is to simply adopt that $\epss$ for a given
%clump takes on the smaller of the two values given by equations
%(\ref{eq:epss}) and (\ref{eq:epss_explosive}).

\smallskip
One can compute clump depletion lifetimes in an analogous manner. 
In the case of depletion by a steady wind, the lifetime is 
\be
\tdep = \frac{M_\star}{\SFR}
= \frac{\Mc}{\Mdots + \Mdotw} 
%\simeq  2 \frac{\epss}{\epsf} \tff \, ,
\simeq  \frac{2}{\epsf(1+\eta)} \tff \, ,
\label{eq:tdep}
\ee
where $\Ms=\epss\Mc$ is the final stellar mass,
%\adb{
with $\epss$ from \equ{epss}.
%}
%and we have assumed in the last equality $\Mg \simeq 0.5 \Mc$ in the 
%expression for the SFR.
%\adb{
For the SFR that enters the last equality of \equ{tdep} we have replaced $\Mc$
in \equ{sfr} by $0.5 \Mc$, to represent the average between the initial
gas mass of $\Mc$ and the final zero gas mass at depletion, 
and thus refer to the characteristic SFR during the period from the onset of
the wind to depletion.
%}
With $\eta \sim 1$, this is much larger than $\tff$.
In the case of explosive disruption the depletion time 
%\adb{
of the explosive phase
%}
is simply of order $\tff$. 

%------------------------------------------
\subsection{Clump Migration versus Depletion}

% migration
In the case of giant clumps in high redshift galaxies, the lifetime may 
also be limited because after some period a clump will \textit{migrate} 
into the 
galactic centre following angular momentum and energy loss by torques from
the perturbed disc, clump-clump interaction and dynamical fraction. 
The time required for this to happen is \citep{dsc09, cdb10}
\be
\tmig \simeq 2.1 Q^2 \delta^{-2} \td \simeq 8\,\td,
\label{eq:tmig1}
\ee
where
\be
\td = \frac{\Rd}{\Vd} 
\ee
is the disc crossing time, $\Rd$ is the characteristic
disc radius and $\Vd$ is the characteristic disc circular velocity.
The quantity $\delta$ is the mass fraction in cold disc
within the disc radius, which at the cosmological steady state is $\delta
\simeq 0.33$. The Toomre parameter is $Q \sim 0.68$ for
a thick disc \citep{goldreich65_thick,dsc09}. 
The migration time is thus comparable to the orbital time at the outer disc.
If we approximate $\td \simeq 3 \tff$, assuming that the clumps are
overdensities of $\sim 10$ with respect to the background disc
\citep[e.g.][]{ceverino12}, we get
\be
\tmig \simeq 24 \tff \simeq 12\, \epsf (1+\eta)\, \tdep \, .
\label{eq:tmig-tdep}
\ee
The true clump lifetime will be the lesser of $\tdep$ and $\tmig$, 
and the corresponding star formation efficiency will be the lesser of
$\epss$ and 
\be
\epsilon_{*,\rm mig} \simeq \frac{\SFR}{\Mc} \tmig 
= \epsf \frac{\tmig}{\tff} \simeq 24\,\epsf \, ,
\ee
where the last equality assumes again $\td \simeq 3 \tff$.
This expression for $\epsilon_{*,\rm mig}$ is valid when $\tmig$ is 
significantly smaller than $\tdep$,
so the approximation $\Mg \sim \Mc$ in the SFR is good. 
If $\tmig$ and $\tdep$ are comparable,
then a better approximation for $\epsilon_{*,\rm mig}$ should be smaller 
by a factor of $\sim 2$. 
From \equ{tmig-tdep}, if $\epsf \simeq 0.01$, we learn that $\tmig$ is 
expected to be
smaller than $\tdep$ as long as $(1+\eta) < 8.3$.
When $\tmig$ is the shorter timescale, namely when $\eta$ is of order unity,
the clump bound mass fraction remaining after outflow mass loss at the end 
of the migration is
\be
\frac{\Mc_{,\rm mig}}{\Mc} \simeq 1 -\frac{\Mdotw \tmig}{\Mc}
\simeq 1 - \eta\, \epss_{,{\rm mig}} \, .
\ee
For $\eta \sim 1$ this is a significant fraction of the original clump mass.

%-----------------------------
\subsection{Clump Mass Growth during Migration}
\label{sec:acc}

% entering the tidal radius 
Accretion onto the clumps, including clump mergers
and possibly tidal stripping of the clumps, are significant during
the clump migration inward, and may actually be the dominant effect in the
evolution of clump mass.
As the clump spirals in toward the disc centre it accretes matter from the
underlying disc. An estimate of the accretion rate is provided by the
entry rate into the tidal (Hill) sphere of the clump in the galaxy, $\Rt$,
\be
\dot{M}_{\rm ac} \simeq \alpha\, \rho_{\rm d}\,(\pi \Rt^2)\,\sigma_{\rm d} \, .
\label{eq:Mdot_ac}
\ee
Here $\rho_{\rm d}$ is the density in the cold disc (gas or young stars),
$\pi\Rt^2$ is the cross section for entry into the tidal sphere,
and $\sigma_{\rm d}$ is the velocity dispersion in the disc representing
here the relative velocity of the clump with respect of the rest of the disc.
The parameter $\alpha$ is expected to be of order unity and smaller.

% tidal radius
\smallskip
The tidal or Hill radius $\Rt$ about the clump is where the self-gravity force
by the clump balances the tidal force exerted by the total mass distribution
in the galaxy along the galactic radial direction.
If the disc is in marginal Toomre instability with $Q \sim 1$,
this is the same as the Toomre radius of the proto-clump patch
that contracts to form the clump \citep{dsc09},
\be
\Rt \simeq 0.5\,\delta \Rd\, ,
\label{eq:Rt}
\ee
where the clump mass is given by
\be
\frac{\Mc}{\Md} \simeq \left( \frac{\Rt}{\Rd} \right)^2 \, ,
\label{eq:Mc}
\ee
with $\Md$ referring to the mass of the cold disc.
%(one does not need here the explicit expression for $\Rt$).
Also when $Q \sim 1$, the disc half height $h$ is comparable to $\Rt$,
\be
\frac{h}{\Rd} \simeq \frac{\sigma_{\rm d}}{\Vd} \, ,
\label{eq:hr}
\ee
%\adb{
and  
\be
\delta \simeq \sqrt{2}\frac{\sigd}{\Vd} \, .
\label{eq:delta-sv}
\ee
%}

%\smallskip
%\adr{The following paragraph was brought here from 3 paragraphs below}

%tac
\smallskip
%\adb{
We can now evaluate the timescale for clump growth by accretion, $\tac$,
using \equ{Mdot_ac}. We insert $\Rt$ from \equ{Rt},
write $\rhod = \Md/(2\pi\Rd^2 h)$, and use \equ{hr} for $h$ to obtain
%}
%The above put together yield for the timescale for clump growth by accretion
\be
\tac \equiv \frac{\Mc}{\dot{M}_{\rm ac}} \simeq \frac{2}{\alpha}\td \, .
\label{eq:tac}
\ee
With $\alpha \sim 1/3$ 
%\adb{
(see below),
%}
the timescale for doubling the clump mass by accretion
is roughly an orbital time, comparable to the migration time.

% binding new
\smallskip
The parameter $\alpha$ represents the fraction of the mass entering the tidal
radius that is actually bound to the clump. If the clump collapses from an
initial patch of radius $\Rt$, the particles enter the tidal
radius with a velocity distribution that is similar to that of the overall 
disc, namely
with a standard deviation $\sigd$, and a distribution of kinetic energies
per unit mass about $0.5\,\sigd^2$ in the clump rest frame.
%\adb{
Using \equ{Rt}, \equ{Mc} and \equ{delta-sv},
%} 
the binding potential of the clump at $\Rt$ can be crudely estimated by 
\be
\frac{G\Mc}{\Rt} \sim \sigd^2 \, , 
\label{eq:pot_Rt}
\ee
so a significant fraction of the
particles entering the tidal radius are expected to be bound.

\smallskip
One can estimate $\alpha$ by referring to the particles that
actually hit the clump, of radius $\Rc < \Rt$, and are bound there.
This requirement puts an upper limit on the
particle impact parameter $b$ prior to entering $\Rt$ such that the focusing
of the orbit would bring the particle into $\Rc$ with a velocity smaller 
than the escape velocity from the clump $\Vc$ at $\Rc$.
Angular-momentum conservation yields $b \simeq \Rc \Vc / \sigd$.
%\adb{
Using \equ{tff} for $\Vc$ and \equ{pot_Rt} for $\sigd$ we obtain
$\Vc/\sigd \simeq (\Rt/\Rc)^{1/2}$.
%}
Therefore in \equ{Mdot_ac}
\be
\alpha \simeq b^2/\Rt^2 \simeq \Rc/\Rt \, .
\ee
With a typical contraction factor of $\Rt/\Rc \simeq 3$ \citep{ceverino12},
the estimate is $\alpha \simeq 1/3$.

% exp disc
\smallskip
For a uniform disc, the relevant density $\rho_{\rm d}$ is the mean density
in the disc, $\bar\rho_{\rm d}$, within a cylinder of radius $\Rd$ and height 
$2h$, and $\alpha$ is the same throughout the disc.
However,
for an exponential disc with an exponential radius $\Rexp$, the local
density $\rho_{\rm d}$ is $\simeq (0.28, 0.46, 0.70)\,\bar\rho_{\rm d}$ 
at $(3,2,1)\,\Rexp$, respectively.
%\adb{
Therefore, if one uses the mean density in \equ{Mdot_ac},
the effective value of $\alpha$ in the outer disc 
could be slightly smaller than estimated above.
%}
%Therefore,  the value of $\alpha$ in the outer disc
%could be slightly smaller than estimated above.

%mergers
\smallskip
The clump growth rate is further enhanced by mergers of clumps as they
spiral in.
If the clumps contain 20\% of the disc mass \citep{dsc09},
and if we require binding when the clump centres are at a distance
of $2\Rc$ from each other, an analogous estimate to \equ{Mdot_ac}
gives that the timescale for growth by mergers is roughly
$t_{\rm mer} \simeq (5/2) \tac$.
On the other hand, tidal stripping may become more pronounced at
small radii, which may slow down the mass growth rate at the late stages of the
migration. We expect the timescale for mass growth to be comparable to
the crude estimate in \equ{tac}.

% mass growth in sims
\smallskip
The above estimates for $\tac$ and $t_{\rm mer}$
are indeed consistent with the findings from
hydro-cosmological simulations, in which outflows by stellar feedback
are weak by construction.
In these simulations, the clump mass is found to be roughly
inversely proportional to distance from the disc centre \citep{mandelker13}.
When following individual clumps as they accrete, strip and merge during
migration, they indeed grow in mass on a timescale that is comparable to the
migration timescale. An effective value of $\alpha \sim 0.33$ in \equ{tac}
seems to provide a good fit to the overall clump mass growth rate in these
simulations (with negligible outflows).

%%%%%%%%%%%%%%%%%%%%%%%%%%%%%%%%%%%%%%%%%%%%%%%%
\section{Momentum Budget}
\label{sec:budget}

Having developed a basic framework for how the properties of star-forming
clumps -- their star formation efficiencies, lifetimes, and outflows -- depend
on the 
%energy and 
momentum injected by stellar feedback, we now turn to various
possible feedback mechanisms. Our goal is to understand the momentum injection
efficiency $\fin$ for each mechanism. 
%In some cases we will also consider the energy budget $\Psi_E$, 
%defined such that the energy release rate via that mechanism
%is $\dot{E} = \Psi_E \dot{M}_*$. 
%\adr{Do we really refer to the energy budget?}
We note that a similar budgeting exercise has
been carried out by \citet{matzner02} for Galactic giant molecular clouds, and
our general approach will follow his, applied to a quite different context.

\smallskip
Unless otherwise stated, all the values below are derived using 
a \textsc{starburst99} \citep{leitherer99a, vazquez05a} calculation for 
continuous star formation, with all parameters set to their default values
except that we use an IMF upper limit of 120 $\msun$ instead of
the default of 100 $\msun$,
though the difference in most quantities is small ($<10\%$). We use
120 $\msun$ because it is the largest mass for which evolutionary tracks
are available in \textsc{starburst99}, and because observations indicate
that the IMF extends to at least this mass if not significantly higher 
\citep{crowther10a}.
%\adr{Say why 120, and what is the difference re using 100}
We evaluate all quantities
at a time of 100 Myr after the start of star formation, but since luminosity and
all other quantities are slowly varying at times $>4$ Myr (which is the essence
of the old stars limit), different choices of age in the range $10-300$ Myr
make a difference of at most a few tens of percent. Given this uncertainty,
we give all results to two significant digits only.

%-------------------
\subsection{Stellar Radiation}
\label{sec:stellarrad}

Stellar radiation pressure has received a great deal of attention recently as a
potential mechanism for disrupting giant clumps, both in analytic models
\citep{murray10,kd10} and in numerical simulations 
\citep{hopkins11a,hopkins12c,hopkins12b,genel12a}.
As stated above, our \textsc{starburst99} calculation gives $\Vl = 190\kms$
as the momentum budget of the direct radiation field.

\smallskip
%Recall our definition of
%$\ftrap$ in \equ{ftrap} as the ratio of the momentum actually injected 
%into the gas to that which would be imparted by direct radiation pressure 
%alone, minus one. Thus $f_{\rm trap} = 0$ 
Recall our definition of $\fin = 1+\ftrap$ as the ratio of the momentum 
actually injected into the gas to that
which would be imparted by direct radiation pressure alone. Thus $\fin = 1$ 
corresponds to a flow that receives no momentum from any
source but photons that are absorbed once and then escape.
If the radiation emitted by stars is trapped by the high optical depths of a
dust layer, it is possible that the trapped photons will be absorbed or
re-emitted multiple times before escape. If this occurs, radiation energy can
build up, and the adiabatic expansion of the radiation-dominated region can
result in a larger momentum transfer to the gas. Some analytic and numerical
models of radiation pressure-driven feedback assume that this effect will
produce a value of $\ftrap\sim \tau$, where $\tau$ is an approximate
infrared optical depth \citep{murray10, hopkins12b, genel12a}, while others
assume that radiation-driven flows are strictly momentum-limited, with 
%$\ftrap$
$\fin$ never exceeding a few \citep{krumholz09d, fall10a, kd10}, due to
radiation Rayleigh-Taylor instability \citep{jacquet11a}. 
This instability punches
holes in the gas that allow photons to leak out, and prevent the buildup of
adiabatic radiation-dominated regions. 

% Krumholz and Thompson
\smallskip
Recent radiation-hydrodynamic simulations by \citet{krumholz12c, krumholz13}
have significantly clarified the matter. They show that, in the case of a 
radiatively-driven wind, the trapping factor obeys $\ftrap \approx 0.5\tau_*$, 
where $\tau_*$ is the optical depth evaluated using the opacity at the dust 
photosphere, \textit{not} the far higher opacity found deep in the dust gas 
where the radiation temperature is higher, as assumed e.g.~by 
\citet{hopkins12b}.
In the old stars limit, for an object of SFR per unit area 
$\dot{\Sigma}_*$
and gas surface density $\Sigma_{\rm gas}$ this is given by \citep{krumholz13}
\be
\tau_* = 0.01\, (c\Vl)_{10}^{1/2}\, \dot{\Sigma}_{*,0}^{1/2}\,
 \Sigma_{\rm gas,0} \, ,
\ee
where $(c\Vl)_{10} = (c\Vl)/(10^{10}\,\lsun/(\msun\mbox{ yr}^{-1}))$, 
$\dot{\Sigma}_{*,0} = \dot{\Sigma}_*/(1\,\msun\mbox{ yr}^{-1}\mbox{ kpc}^{-2})$,
and $\Sigma_{\rm gas,0} = \Sigma_{\rm gas}/(1\mbox{ g cm}^{-2})$.
The values of $\dot{\Sigma}_*$ and $\Sigma_{\rm gas}$ to which we have
scaled are typical of observed giant clumps, as discussed below, and thus
we typically have $\tau_* \ll 1$, and therefore $f_{\rm trap,rad} \approx 0$.
We conclude that the very large trapping factors of $10-50$ assumed 
in certain simulations \citep[e.g.][]{oppenheimer08,genel12a,hopkins12b} 
are unrealistic.

\smallskip
Note that the energy in the wind is 
\be
\dot{E}_{\rm w} = \frac{1}{2} \Mdotw \Vw^2  = \frac{1}{2} \fw \frac{\Vw}{c} L
\, .
\ee
Thus, as long as $\fw$ is of order a few and $\Vw \ll c$, only a small friction
of the photon energy is used to drive the outflow. 
This is indeed 
momentum-conserving rather than energy-conserving driving of the outflow.

%------------------
\subsection{Photoionized Gas}

In Galactic giant molecular clouds, the pressure of photoionized gas is likely
the dominant feedback mechanism that limits the star formation efficiency
\citep[e.g.][]{whitworth79, williams97a, matzner02, krumholz06d, goldbaum11a}.
Photoionization raises gas to a nearly fixed temperature of roughly $10^4$ K,
and that temperature is maintained by radiative heating and cooling processes.
As the gas expands, it transfers momentum to the surrounding medium. By
integrating over the IMF and using a similarity solution to compute the
evolution of expanding H~\textsc{ii} regions, \citet{matzner02} estimates that
this mechanism injects momentum at a rate $\Vin \simeq 260 \kms$.
While this is probably the dominant feedback mechanism for GMCs
in local galaxies, it is likely to be unimportant for
giant clumps, for the simple reason that such clumps have characteristic speeds
significantly higher than the ionized gas sound speed, 
$c_{\rm i} \simeq 10 \kms$ \citep{kd10}.  
As a result, ionized gas will be unable to expand and
transfer momentum to the cold gas. Recent numerical simulations confirm this
conjecture \citep{dale12a}. We may therefore disregard this mechanism for giant
clumps at high redshift. 

%--------------------------
\subsection{Protostellar Outflows}

Protostars drive collimated hydromagnetic outflows with launch speeds
comparable to the escape velocity from stellar surfaces, typically 
$\sim 100\kms$
for protostars with radii larger than those of main sequence stars.
The wind material shocks against and mixes with the surrounding dense molecular
gas. Because the environment the winds encounter is very dense, and the shock
velocity is not high enough to heat material past the $\sim 10^5$ K peak of the
cooling curve, the post-shock gas rapidly cools via radiation, so there is no
significant adiabatic expansion phase. \textsc{Starburst99} does not include
protostellar winds, so we adopt an estimate of
\be
\Vps \simeq 40 \kms \, , 
\ee
from \citet{matzner02}.

%--------------------
\subsection{Supernovae}

From our \textsc{starburst99} calculation, supernovae occur at a rate 
%\adb{
\be
\tau_{\rm sn}^{-1} = 0.012\,(\dot{M}_*/M_\odot\mbox{ yr}^{-1}) \, ,
\label{eq:tsn}
\ee
%}
%\adr{Is this from starburst99?}
carry an energy\footnote{Note that $V$ and $V^2$ here are the energy and momentum
per unit mass of stars formed, not per unit mass of stars that actually end
their lives as SNe.}
$V^2 \simeq 5.8\times 10^{15}$ erg g$^{-1}$,
%\adr{This is roughly $10^{50}$ erg for a $10\msun$ star. Explain how this is
%related to the $10^{51}$ erg assumed for a SN, e.g. in footnote 5.} 
and carry a momentum flux that corresponds to
\footnote{The 
publicly-available version of 
\textsc{starburst99} does not calculate the supernova momentum flux. 
We have modified it to do so, using the same assumptions \textsc{starburst99}
adopts in order to compute the supernova energy injection rate and mass
return, i.e.~all stars with an initial mass above 8 $\msun$ end their lives as
supernovae with identical energies of $10^{51}$ erg. They all leave behind
as remnants $1.4$ $\msun$ neutron stars, so the mass of the ejecta
is simply the final stellar mass (smaller than the initial mass due to wind 
losses) minus the remnant neutron star mass.
}
\be
\label{eq:vsn}
\Vsn \simeq 48 \kms \, . 
\ee
This represents only a lower limit on the true momentum injected by supernovae,
because the extremely high post-shock temperatures produced when supernova
ejecta encounter the ISM guarantee that radiative cooling is inefficient at 
the early stages.
As a result, supernova remnants experience an energy-conserving phase when they
are small, then begin to cool radiatively only once adiabatic expansion 
lowers their internal temperatures sufficiently. During the
adiabatic Sedov-Taylor phase and the pressure-driven snowplow phase that 
follows it (during which the remnant interior is partially radiative), 
the radial momentum carried
by the swept-up material increases. This process has been studied by numerous
authors \citep[e.g.][]{chevalier74a, mckee77a, cioffi88a, thornton98a}, 
and for a uniform medium \citeauthor{thornton98a}~find that the 
asymptotic momentum of a supernova remnant is
\be
%p_{\rm sn}=3\times 10^{43} E_{51}^{13/14}\, \n0^{-0.25}\mbox{ g cm s}^{-1}\, ,
p_{\rm sn}=1.7\times 10^{43} E_{51}^{13/14}\, \n1^{-0.25}\mbox{ g cm s}^{-1}\, ,
\label{eq:psn}
\ee
where $E_{51}$ is the energy of the supernova in units of $10^{51}$ erg 
and 
%$\n0$
$\n1$ is the ambient number density in units of 
%$1$
$10$ cm$^{-3}$. A simple
estimate for the supernova momentum budget including the adiabatic phase
is simply 
%\adb{
obtained from $\dot{p}_{\rm sn}$, which is given by $p_{\rm sn}$ from \equ{psn}
multiplied by the supernova rate $\tau_{\rm sn}^{-1}$ from \equ{tsn}. This gives
%}
\be
\label{eq:vsnadiab}
%V_{\rm sn, adiab} = 1900\, \n0^{-0.25}\mbox{ km s}^{-1} \, ,
V_{\rm sn, adiab} = 1100\, \n1^{-0.25}\mbox{ km s}^{-1} \, ,
\ee
assuming the energy of a single supernova is $E_{51}=1$ (as assumed in the
\textsc{starburst99} calculation as well). Since the momentum input from a
single supernova remnant is very close to linear in $E_{51}$, the total
momentum budget is not significantly affected by the manner in which
the supernovae are clustered.

\smallskip
However, we caution that this calculation is for a simple, one-dimensional
uniform medium. As the \citet{krumholz12c, krumholz13}
results for radiation pressure show, this assumption can be deeply
misleading about how effectively energy is converted into momentum
in a real three-dimensional medium where instabilities can occur. 
It is therefore best to regard eq.~(\ref{eq:vsnadiab}) as representing an 
upper limit.  Determining where reality lies in between this value and the 
lower limit represented by
eq.~(\ref{eq:vsn}) requires numerical simulations capable of following
instabilities into the non-linear phase. Although such simulations have begun
to appear in the literature \citep{creasey13a}, the problem remains far
from fully solved.

\subsection{Main Sequence and Post-Main Sequence Stellar Winds}

The winds of main sequence and post-main sequence stars carry an energy
and momentum content $V^2 \simeq 1.5\times 10^{15}$ erg g$^{-1}$ and
\begin{equation}
V_{\rm ms,dir} \simeq 140 \kms,
\end{equation}
respectively. While they therefore carry slightly less
momentum than the stellar radiation field, at least some of the winds are 
launched
at velocities large enough that the post-shock gas may have long cooling times.
(This is in contrast to the much slower protostellar outflows.) As a result,
it is plausible that stellar winds
could experience an adiabatic phase like supernovae, and enhance their momentum
transfer that way. In this case the momentum provided by winds will 
have an additional term that we can write as $f_{\rm ad} \Vl$.
%\be
%\Vms = \Psi_{p,\rm wind} + f_{\rm trap,wind} \Psi_{p,\rm rad}
%\ee

\smallskip
The main idea of the classical stellar wind bubble model of \citet{castor75a}
and \citet{weaver77a} is that $f_{\rm ad} \Vl \gg V_{\rm ms,dir}$.
On the other hand, this mechanism will not operate if
wind gas is able escape from a star-forming clump without entraining
significant mass, or if it undergoes rapid cooling by mixing with cooler, dense
gas that brings its temperature low enough for radiative losses to become
rapid. 
Observations of a few nearby H~\textsc{ii} regions have been able to address
this question directly by using x-ray observations to probe the energy density
of the shock-heated gas. Both \citet{harper-clark09a}, who study the Carina
Nebula, and \citet{lopez11a}, who study 30 Doradus, find that the luminosity of
the x-ray emitting gas implies that the pressure exerted by this gas is weaker
than that exerted by photoionized gas, a result highly inconsistent with an
energy-driven flow.\footnote{Contrary to the results of \citet{lopez11a},
\citet{pellegrini11a} argue that the x-ray emitting gas pressure in 30 Doradus
is actually higher than the ionized gas pressure. \citeauthor{pellegrini11a}'s results differ because
they assume that the x-ray emitting gas is confined to a small volume. Since
the x-ray luminosity is proportional to the emission measure of the emitting
gas, which is the integral of the square of the electron density along the line
of sight, if one assumes that the line of sight length is much smaller than the
transverse size of the region being observed, the density and thus pressure
that one infers for a given observed luminosity rises proportionately. For our
purposes, however, this distinction is irrelevant. If one assumes that the hot
gas is confined to a small fraction $f_X$ of the observed volume, the pressure
$P_X$ inferred from a given luminosity varies as 
$P_X\propto f_X^{-1/2}$, but the
energy content of the hot gas, which varies $f_X P_X$, falls as $f_X^{1/2}$.
Thus if \citeauthor{pellegrini11a}'s conjecture about the geometry is correct,
that implies even more strongly that feedback from stellar winds cannot be
energy-driven.} 

\smallskip
For our fiducial estimate in this paper we adopt Lopez et al.'s measured mean
value $f_{\rm ad} = 0.3$, so that the net amount of momentum ejected by
the winds from main-sequence stars is 
\be
\Vms = V_{\rm ms,dir} + f_{\rm ad} V_{\rm L} \simeq 200\kms \, .
\ee
%\adr{Check this number.}
However, we caution that none of the observed regions have conditions close 
to those
of high redshift giant clumps. While stellar wind gas is momentum- and not
energy-driven in the local universe, a giant clump could be considerably harder
for hot x-ray gas to escape. It is therefore conceivable that, under the
conditions found in high redshift giant clumps, stellar winds represent an
energy-driven feedback. In this case it is likely that stellar wind and 
supernova bubbles would simply add together to produce an adiabatic shell 
driven by the combined effects of both. The result is to increase the 
adiabatic energy budget by roughly 25\% compared to supernovae alone. 
Adopting this simple estimate, we find that if winds are adiabatic then, 
in conjunction with supernovae, the net momentum contribution of the winds is
\be
%V_{\rm ms,adiab} = 490\, \n0^{-0.25}\mbox{ km s}^{-1}.
V_{\rm ms,adiab} = 275\, \n1^{-0.25}\mbox{ km s}^{-1}.
\ee
Note that this estimate implicitly assumes that the temperature and cooling are
determined by the significantly larger energy associated with the supernovae, 
so that stellar winds simply pump more energy into the adiabatic bubble without
significantly affecting how it cools.

\subsection{Total Momentum Budget}

Combining all the mechanisms we have enumerated (excluding photoionized gas for
the reasons stated above), we see that the momentum budget
for the case of a purely momentum-driven outflow is expected to be 
\be
\Vin = \Vrad + \Vps + \Vsn + \Vms \simeq 480 \kms \, .
\label{eq:Vin_min}
\ee
The corresponding trapping factor is rather small, 
\be
\fin = 1+\ftrap \simeq 2.5 \, . 
\label{eq:fin_min}
\ee
By ``purely momentum-driven outflow" we refer to the case where radial momentum
is conserved and neither supernovae nor stellar winds experience an 
energy-conserving phase during which their radial momentum is significantly 
boosted. 
This can be interpreted as a lower limit for $\fin$ from stellar feedback.

\smallskip
If we assume that there is a significant adiabatic phase for supernovae and 
main-sequence winds, and that the resulting momentum injection is near the 
upper limit derived in the uniform medium case, then we obtain an upper limit
for the net momentum budget of
\begin{eqnarray}
V_{\rm in,adiab} & = & \Vrad + \Vps + V_{\rm sn,adiab} + V_{\rm ms, adiab}
\nonumber \\
%& \simeq & 280 + 2400\,\n0^{-0.25} \kms\, ,
& \simeq & 230 + 1350\,\n1^{-0.25} \kms\, ,
\label{eq:Vin_max}
\end{eqnarray}
%\adr{Check the 280 -- shouldn't it be 230?}
%corresponding to an upper-limit value of
\be
%\fin \simeq 1.5+12.6\,\n0^{-0.25} \, .
\fin \simeq 1.2+7.1\,\n1^{-0.25} \, .
\label{eq:fin_max}
\ee
For massive clumps of $\tff \simeq 7\Myr$ this is $\fin \simeq 6.4$.

%\medskip
%\adr{
%Use $n_{10}$ or $n_{20}$?
%$n=22 t_{10}^{-2} \gcmc$. If $\tff=7$ then a typical clump is $n = 44$.
%$\Sigma = 1 \gcms = 4500 \msun \pc^{-2}$.
%In a typical clump $\Sigma = 0.1-1\gcms$.
%}

\smallskip
The actual contribution of adiabatic supernova feedback and the corresponding
value of $\fin$ between the above lower and upper limits is a matter
of a saturated state of fully nonlinear instabilities, which should be
determined by appropriate numerical simulations.
The two most relevant publications to date on this subject are
\citet{hopkins12b} and \citet{creasey13a}. For the former, if we
examine the runs without the subgrid radiation model where supernova
feedback dominates, the mass loading factors are $\eta \sim 1-5$ and the
wind terminal speeds are a few hundred km s$^{-1}$, implying $\fw$ values
of a few; combined these suggest $\fin \sim 5$. Similarly, \citet{creasey13a}
report a mass loading factor $\eta$
and a wind thermalization parameter $\eta_T$ (which measures the fraction
of supernova energy that goes into outflow; see their equation 5), both 
as a function of galaxy properties. With some algebra, one can show that
$\fw = 5.3 (\eta\eta_T)^{1/2}$, and using their fitting formulae for galaxy
surface densities $\sim 10-100$ $\msun$ pc$^{-2}$, appropriate to giant
clump galaxies, gives $\fw \sim 1-3$ for supernovae alone. In summary, the
numerical results of both \citet{hopkins12b} and \citet{creasey13a} suggest
that $\fin$ is likely to be roughly halfway between our upper and lower limits.

\smallskip
A potential way to make the supernova feedback more effective is by having 
the gas density in the supernova vicinity much lower than the unperturbed
density within the clump of $\n1 \geq 1$. 
A value of $\n1 \sim 10^{-2}$ (or $10^{-3}$) in \equ{fin_max} 
would provide a maximum value of $\fin \simeq 24$ (or 41 respectively).
The question is whether the other types of momentum-conserving stellar 
feedback could generate such a low-density regime prior to the supernova 
explosion.  We keep this mechanism outside the scope of the present paper.

%%%%%%%%%%%%%%%%%%%%%%%
\section{Implications for High-Redshift Giant Clumps}
\label{sec:implications}

%---------------------------
\subsection{Star Formation and Outflows}

In the preceding two sections, we developed a general framework
to consider the evolution of clumps as they migrate, accrete, form stars,
and lose mass due to star formation feedback, and we derived an estimate
for the momentum budget of the feedback that drives clump winds. We
now combine these results to draw conclusions about the typical
evolutionary path taken by giant clumps.

\smallskip
The structure and dynamics of a clump are characterized by two quantities,
e.g., its characteristic velocity and its free-fall time, 
$\Vc \equiv 100 \kms V_2$,
and 
$\tff \equiv 10 \Myr\, \tff_{10}$,
as defined in \equ{tff}.\footnote{The relations to the clump mass and radius
are
$V_2 \simeq 1.15 M_{9.5}^{1/2} R_1^{-1/2}$
and
$t_{10} \simeq 0.96 R_1 V_2^{-1} 
\simeq 0.82 R_1^{3/2} M_{9.5}^{-1/2}
\simeq 1.27 M_{9.5} V_2^{-3}$
where $M_{9.5} \equiv \Mc/10^{9.5} \msun$ and $R_1 \equiv \Rc/1 \kpc$.
Also $\n1 \simeq 2.2 t_{10}^{-2}$.
The surface density is $\Sigma \simeq 0.21 M_{9.5} R_1^{-2} \gcms$ 
and $1 \gcms \simeq 4800 \msun\pc^{-2}$.}
 
\smallskip
Following the earlier discussion,
the physics of outflow from giant clumps can be characterized by three 
dimensionless parameters, e.g., $\epsf$, $\nu$ and $\fw = \fej \fin$.
These are the SFR efficiency $\epsf \equiv 0.01 \epsf_{,-2}$,
the wind velocity with respect to the clump escape velocity,
$\nu \equiv \Vw/\Vc \equiv 3\nu_3$,
and the wind momentum with respect to the radiation momentum,
$\fw = \pdotw/(L/c)$.
%\adb{
For the values of $\epsf$ and $\fin$ we have theoretical predictions. 
For the values of $\nu$ and $\fej$, unity is a lower and an upper limit
respectively, but we do not have a theoretical prediction concerning how
much they actually deviate from unity.
Motivated by observations (see below), we assume that these deviations
are by a multiplicative factor of order one or a few.
%}
We define 
$\fw \equiv 2.5 \fw_{,2.5}$. 
The reference value of $\fw \simeq  2.5$ may refer to the case of pure 
momentum-driven outflow $\fin \simeq 2.5$ and maximum ejection of 
$\fej \simeq 1$,
or to a case including adiabatic supernova and stellar-wind feedback
with $\fin \simeq 5$ but with some losses in the ejection, $\fej \simeq 0.5$.
The maximum value of $\fw$, when adiabatic supernova feedback is at its maximum
and the ejection is efficient, is expected to be $\fw \sim 5$.  

%$\fej \equiv 0.5 \fej_{,0.5}$, and
%$\fin \equiv 5 \fin_{,5}$.

\smallskip
The SFR and wind mass flow rate are
\be
\SFR \simeq 2.4 \epsf_{,-2} V_2^3 \sy \, ,
\label{eq:sfr2}
\ee
%\adr{check the number above}
\be
\Mdotw \simeq 3.2 \epsf_{,-2} {\psi}_{\rm w,2.5} \nu_3^{-1} V_2^2 \sy \, .
\label{eq:Mdotw2}
\ee
%\adr{check the number above}
The corresponding timescales are
\be
\tsfr = \frac{\Mc}{\SFR} \simeq 1\Gyr\, \epsilon_{\rm ff,-2}^{-1} \tff_{,10} \, ,
\label{eq:tsfr2}
\ee
\be
\tw = \frac{\Mc}{\Mdotw} 
\simeq 1 \Gyr\, \fw_{,2.5}^{-1} \Vw_{,400} \epsilon_{ff,-2}^{-1} \tff_{,10} \, .
\label{eq:tw2}
\ee
The mass loading factor, \equ{eta}, is
\be
\eta \simeq {\fw}_{,2.5} {\Vw}_{,400}^{-1} 
= 1.33 {\fw}_{,2.5} \nu_3^{-1} V_2^{-1} \, ,
\label{eq:eta2}
\ee
where $\Vw \equiv 400\kms {\Vw}_{,400}$.
With the fiducial values adopted here for momentum-driven stellar feedback
from typical clumps
one expects steady winds with mass-loading factors of order unity.
The maximum value, when adiabatic supernova feedback is included, is
expected to be $\eta \sim 2-3$.
 
\smallskip
If the clump were allowed to deplete all its gas,
the final star formation efficiency would have been 
\be
\epss = (1+\eta)^{-1} \equiv 0.5 (1+\eta)_2^{-1}
\label{eq:epss2}
\ee
at the clump depletion time of
\be
\tdep = 1 \Gyr\, (1+\eta)_2^{-1} \epsf_{,-2}^{-1} \tff_{,10} \, ,
\label{eq:tdep2}
\ee
where $(1+\eta)_2 \equiv (1+\eta)/2$.

\smallskip
The approximate values for $\epss$ and $\tdep$ are valid  
when $\epss$ deviates significantly from unity, and where clump disruption 
is by gradual erosion rather than sudden explosive destruction. 
%The threshold efficiency for explosive ejection, \equ{epssexp}, is
%\be
%\epss_{,\rm exp} \simeq 1- 0.02 \epsf_{,-2} \fw_{,2.5} V_2^{-1} \, .
%\ee
%\mkb{
The criterion for explosive disruption, \equ{explosive},
is simply
\be
\epsf_{,-2} \fw_{,2.5} V_2^{-1} \gsim 20.
\ee
%}
This is similar to eq.~9 of \citet{kd10}, where the considerations were
qualitatively similar though not exactly the same numerically.
As noted there, if the clump is a typical Toomre clump
with $\Vc \sim 100\kms$, 
%$\epss_{,\rm exp}$,
%\mkb{
explosive disruption occurs
%}
%could be significantly smaller than unity
only if either $\epsf$ or $\fw$ are significantly larger than their 
fiducial values, namely either the SFR is much more efficient than implied
by the local Kennicutt relation, or the momentum-driven feedback is 
much more efficient than available in the momentum budget evaluated above.
Otherwise, with the adopted fiducial values for these quantities, 
%\mkb{
explosive disruption does not occur,
%}
%$\epss_{,\rm exp}$ is indeed close to unity and significantly larger than 
%$\epss$
thus validating the steady-wind approximation used.

%--------------------------
\subsection{Clump Migration}

In comparison, the clump migration time is 
\be
\tmig \simeq 8\td \simeq 260 \Myr\, \Rd_{,7} \Vd_{,200}^{-1} \, ,
\ee
where the disc is characterized by $\td = \Rd/\Vd$ with 
$\Rd_{,7} \equiv \Rd/7 \kpc$ and $\Vd_{,200} \equiv \Vd/200\kms$.
The fiducial values of $\Rd$ and $\Vd$ are deduced from observations at 
$z \sim 2$ \citep{genzel06,genzel08}, but the disc dynamical time
can also be derived from the virial radius and velocity using the virial
relation and the spherical collapse model and assuming a constant
spin parameter $\lambda$ for halos and conservation of angular 
momentum during gas collapse within the dark matter halo, 
\be
\td \simeq \lambda \frac{\Rv}{\Vv}
\simeq 0.07\,\lambda_{0.07}\,0.15\,t_{\rm Hubble}\, ,
\ee
which at $z =2$, where $t_{\rm Hubble} \simeq 3.25 \Gyr$,
gives $\td \sim 33\,\lambda_{0.07} \Myr$.

\smallskip
The relation between the depletion time and migration time is
\be
\tmig/\tdep \simeq 0.25\, (1+\eta)_2\, \epsf_{,-2} \, ,
\label{eq:tmig-tdep2}
\ee
where we have assumed $\td \simeq 3 \tff$ for the dynamical timescales
in the disc and in the clumps.
If $\tmig \leq \tdep$,
The maximum star formation efficiency
possible before the clump reaches the galactic centre is
\be
\epss_{,\rm mig} \simeq 0.24 \epsf_{,-2} \, .
\label{eq:epss_mig2}
\ee
The clump bound mass fraction remaining at the end of the migration is
\be
\frac{\Mc_{,\rm mig}}{\Mc} \simeq 1 - \frac{\Mdotw \tmig}{\Mc}
\simeq 1 - 0.24 \eta \epsf_{,-2} \, .
\label{eq:mcmig2}
\ee
These are for $\tmig$ significantly smaller than $\tdep$, namely for $\eta$ 
of order unity.
The estimate for $\epss_{,\rm mig}$ is an overestimate by up to a factor of 
order 2 because we assumed here $\Mg \simeq \Mc$. For the same reason,
the expression for $\Mc_{,\rm mig}/\Mc$ is an underestimate.
With the fiducial values adopted here, the clump reaches the centre
while still holding on to a significant fraction of its original mass,
and most likely still gas rich.

%-------------------------
\subsection{Clump Mass Evolution}

% tac
\Equ{tac}, with $\td \simeq 3 \tff$, yields
\be
\tac \simeq 0.18 \Gyr\, \alpha_{0.33}^{-1}\, \tff_{,10} \, ,
\label{eq:tac2}
\ee
where $\alpha_{0.33}=\alpha/0.33$.
Thus, with $\alpha = 0.33$, 
the timescale for doubling the clump mass by accretion is 
$\sim 6\td \sim 0.18\,\Gyr$,
which is slightly smaller than the migration time, 
$\tmig \sim 8\td \sim 0.24\,\Gyr$. 

\smallskip
With the fiducial values for momentum-driven winds,
the mass growth rate as estimated in \equ{tac2} is faster than the outflow
rate and the SFR, \equ{tw2} and \equ{tsfr2},
implying that the accretion more than compensates for the mass loss by
outflows, making the clumps actually grow in mass as they migrate inwards.
This implies in particular that the adopted estimate of migration time
remains a good approximation and may even be an overestimate.

% growth rate with outflows
\smallskip
The evolution of clump mass $M(t)$ under accretion and outflows,
starting from an original mass $\Mc$ at $t=0$, is governed by
\be
\dot{M} = \dot{M}_{\rm ac} - \Mdotw \, .
\ee
What makes the integration of this equation simple is that the two terms 
on the right hand side both scale with $M/\tff$.
First,
\be
\dot{M}_{\rm ac} \simeq \frac{\alpha}{2} \frac{\tff}{\td} \frac{M}{\tff} \, ,
\ee
where $\alpha$ and $\tff/\td$ are approximated as constants, the latter
being determined by the clump collapse factor from the original protoclump
patch in the disc.
Second,
\be
\Mdotw \simeq \eta\, \epsf \fg \frac{M}{\tff} \, ,
\ee
where $\fg$ is the star-forming gas fraction in the clump, approximated as
constant.
Integrating we obtain
\be
M(t) = \Mc\, {\rm e}^{\gamma t/\tff} \, , \quad
\gamma = 0.5\, \alpha\, (\tff/\td) - \eta\, \epsf \fg \, .
\label{eq:gamma}
\ee
With our fiducial values ($\alpha=0.33$, $\tff/\td=1/3$, $\eta=1$, $\epsf=0.01$,
$\fg=1$) we have $\gamma = 0.045$. 
With $\tmig=8\td$ the growth factor during migration becomes 
$M(\tmig)/\Mc \simeq 2.9$.
It requires a very strong wind of $\eta \sim 5.5$
for the mass loss to 
balance the accretion and leave the clump with a constant mass till depletion, 
which in this case may occur before the clump completes its migration.
%\adb{
For a significant mass loss in a migration time, $\tmig \sim 24 \tff$,
$\gamma$ in \equ{gamma} has to be significantly smaller than $-1/24$.
With the fiducial value of $\alpha=0.33$, this requires that $\eta\epsf$ 
would be larger than its fiducial value of $0.01$ by an order of magnitude.
Alternatively, $\gamma$ could obtain such negative values if the 
effective $\alpha$ is negative, e.g., representing a case where mass loss 
by tidal stripping overwhelms the mass gain by accretion. 
However, the reported significant clump growth in the simulations, where both
accretion and tidal stripping are at play, indicates that the effective
$\alpha$ is positive and close to the assumed fiducial value. 
We conclude that a net mass loss in the clumps is very unlikely.
%}

%----------------
\subsection{Other Implications}

The predictions listed above have a few immediate and interesting implications. 
If winds are relatively efficient, i.e.~$\fej \sim 1$,
then when all types of stellar feedback are taken into account one expects 
giant clumps to experience fairly significant steady winds. Purely
momentum-driven feedback is expected to provide a mass loading
factor $\eta$ of order unity, and adiabatic supernova feedback can boost it
to $\eta$ of a few.
We emphasize that the significant outflows hold even though the
radiative trapping is negligible, and even though the clumps do not experience
explosive disruption in a dynamical time.

\smallskip
If the clumps were allowed to reach depletion, the 
depletion time would have been on the order of a significant fraction of 1 Gyr.
However, the clumps are likely to complete their migration inward at 
a shorter time. During this migration, the clumps accrete mass 
from the disc and merge with other clumps, roughly doubling their mass
in one orbital time.

\smallskip
%The fact that the timescales for star formation and migration are comparable
%and typically shorter than the gas depletion time 
%\adb{
The fact that the timescale for migration is typically shorter than
the timescales for star formation and depletion
%}
indicates that the clumps 
complete their migration while still gas rich, thus taking part in the 
overall ``wet" inflow within the disc \citep{forbes12,cdg12,dekel13}.
This ``wet" inflow has
interesting implications, e.g., it naturally leads to a compact bulge
(Dekel \& Burkert, in prep.) and could feed the central black hole
\citep{bournaud11,bournaud12}.

\smallskip
A question often raised is whether the outflows from clumps can be the driver
of turbulence in the disc, the mechanism that maintains the Toomre instability 
at $Q \sim 1$.
A necessary condition is that the power in the outflows is comparable to the
turbulence dissipative loss.
The outflow power from $N_{\rm c}$ clumps is 
\be
\dot{E}_{\rm w} \sim N_{\rm c} \Mdotw \Vw^2 \, , 
\ee
with $\Mdotw$ and $\Vw$ as predicted above.
The turbulence is expected to decay on a disc dynamical time, so the
dissipation rate is
\be
\dot{E}_{\rm dis} \sim \Mg \sigd^2 / \td \, .
\ee
One can use from our analysis above $\Vw = \nu \Vc$,
$\Mdotw = \eta\sfr$, $\sfr=\epsf \Mg/\tff$.
For a clump contraction factor $c$, the dynamical times are related as
$\td = c \tff$.
For a $Q \sim 1$ disc and clumps one can estimate that the internal clump
velocity and the external disc velocity dispersion are comparable
\citep[e.g.][]{ceverino12}; they are related by
%\adb{
\be
%{\Vc^2}/{\sigd^2} \sim {c}/{4} \, .
{\Vc^2}/{\sigd^2} \sim (\pi/2)\,c \, .
\ee
Then the ratio of
the rate of energy injection by clump winds to
energy loss due to decay of turbulence becomes
\be
\frac{\dot{E}_{\rm w}}{\dot{E}_{\rm dis}} 
%\sim N_{\rm c,5}\, \eta\, \epsf_{-2}\, \nu_3^2\, c_3^2  \, .
\sim 6\, N_{\rm c,5}\, \eta\, \epsf_{-2}\, \nu_3^2\, c_3^2  \, .
\ee 
%}
This seems to indicate that there is enough energy in the outflows to
continuously stir up the disc.
However, it is likely that a large fraction of the outflow energy will
be ejected along the descending density gradient perpendicular to the disc
and not injected into the inter-clump medium in the disc plane,
thus making the contribution of outflows to the disc turbulence only secondary. 
The gravitational gain by the VDI-driven inflow along the potential gradient
within the disc is a more likely source of energy for maintaining the
disc turbulence \citep[e.g.][]{bournaud11,forbes12,cdg12,dekel13}.

%%%%%%%%%%%%%%%%%%%%%%%%%%%%%%%%%%%%%%%%%%%%
\section{Comparison to Observed Clumps}
\label{sec:obs}

\subsection{Observed Clumps}

\begin{table*}
\centering
\caption{Observed properties of giant clumps form \citet{genzel11}. Quantities marked by ``*" are deduced
relatively directly from the observations, while the other quantities are computed from them. }
\begin{minipage}{6.9in}
\centering
\begin{tabular}{lccccccc}
\hline
Clump no.\ \  & 
  1 & 2 & 3 & 4 & 5 & 6 & 7 \\
Clump name \ \  & 
\ \ BX482-A \ \ & D3a15504-A-F & ZC782941-A & ZC406690-C \ \ \ \ & \ \ \ \ ZC406690-A & ZC406690-B  & \ \ \ \ BX599 \ \ \\
$z$* &
      2.3       &  2.4         &  2.2       &   2.2        & 2.2          &  2.2           &    2.3    \\
\hline
$\Rc$ [kpc]* \footnote{$\Rc=R_{\rm HWHM}$ after beam smearing (of HWHM$\simeq 2\kpc$) is subtracted in 
quadrature} &  
      1.0       &  1.0         &   0.8      &   1.2        &  0.8         &   1.2          &    1.5    \\
$\Vc$ [km s$^{-1}$]* \footnote{$\Vc^2 = \beta\, (\Vrot^2 + c\,\sigma^2)$, \ $\beta=1.17$, \ $c=3.4$, \ assuming
Jeans equilibrium} & 
     125        &  111         &   195      &   159        &  163         &    187         &    152    \\
$\Mc$ [$10^9\msun$] \footnote{$\Mc= G^{-1}\Vc^2\Rc$} & 
      3.6       &  2.8         &   6.9      &   6.9        &  4.8         &   9.5          &    7.8    \\
$\tff$ [Myr] \footnote{$\tff = \Rc/\Vc$} & 
      7.7       &  8.7         &   3.9      &   7.2        &  4.7         &   6.2          &    9.5    \\
\hline
$\SFR$ [$M_\odot{\rm yr}^{-1}$]* \footnote{$\SFR=L({\rm H}\alpha)/2.1\times10^{41}\ergs$ extinction corrected}& 
       12       &  3.3         &   17       &   14         &  40        &   11           &    66      \\
$\Vw$ [km s$^{-1}$]* \footnote{$\Vw=\la V_{\rm broad}\ra - 2\,\sigma_{\rm broad}$} &  
      350       &  400         &   420      &   355        &  440       &   810          &   1000     \\
$\Mdotw$ [$M_\odot {\rm yr}^{-1}$]* \footnote{Average of two photodissociation case-B models. Clumps 5 and 6 are from
\citet{newman12}} &  
      12        &  3.6         &   34       &    13        &  117       &   78           &   185      \\
\hline
$\epsf_{,-2}$ \footnote{$\epsf_{,-2}=4.2\,\Vc_{,2}^{-3}\dot{M}_{\star,10}$ (\equnp{sfr2}) } &  
      2.5       &  1.0         &   1.0      &   1.5        &  3.8       &   0.7           & {\bf 7.9}  \\
$\nu$ \footnote{$\nu = \Vw/\Vc$} &  
      2.8       &  3.6         &   2.2      &   2.2        &  2.7       &   4.3          &  {\bf 6.6} \\
$\eta$ \footnote{$\eta = \Mdotw/\sfr$} &  
      1.0       &  1.1         &   2.0      &   0.9        &  2.9       &  {\bf 7.1}     &  2.8       \\
$\fw$ \footnote{$\fw=\Mdotw\Vw/(L/c)$} & 
       2        &  3           &   4        &   2          &  {\bf 6}   & {\bf 34}       & {\bf 14}   \\
%$\fw$\footnote{$\fw=2.5\,\eta\,\Vw_{,400}$ \adr{This row will be removed. I am not sure why these $\fw$ differ. Maybe extinction.}} &  
%      2.2       &  2.7         &   5.2      &   2.1        &  {\bf 8.5} & {\bf 42}       & {\bf 17 } \\
$\Sigma_{\rm gas}$ [g cm$^{-2}$] \footnote{$\Sigma_{\rm gas}=\Mg/(\pi \Rc^2)$, \ $\Mg=\epsf^{-1} \sfr \tff$} &
      0.26      &   0.20       &   0.77     &   0.34       &   0.54     &     0.48       &    0.25   \\
\hline
$\tdep$ [Myr] \footnote{$\tdep=1000\Myr\, (1+\eta)_2^{-1} \epsf_{,-2}^{-1}\, \tff_{,10}$ (\equnp{tdep2})} &
      302       &  817         &   274      &  517         &  {\bf 62}  &  218           & {\bf 63}   \\
$\epss$ \footnote{$\epss=(1+\eta)^{-1}$ (\equnp{epss2}, relevant when $\tmig/\tdep >1$)} &  
     0.50       &  0.48        &   0.33     &   0.52       &  0.25      &   0.12         &    0.26    \\
$\tmig/\tdep$ \footnote{$\tmig/\tdep = 0.25\, (1+\eta)_2\, \epsf_{,-2}$ (\equnp{tmig-tdep2})} &
     0.63       &  0.27        &   0.36     &   0.35       &  1.9       &   0.71         &    3.7     \\
$\epss_{,\rm mig}$ \footnote{$\epss_{,\rm mig}=0.24\, \epsf_{,-2}$ (\equnp{epss_mig2}, relevant when $\tmig/\tdep <1$)} &
  $\sim 0.50$    &  0.24        &   0.23     &   0.35       &  -         &   $\sim 0.12$   &     -      \\
$\Mc_{,\rm mig}/\Mc$ \footnote{$\Mc_{,\rm mig}/\Mc=1-\eta\,\epss_{,\rm mig}$ (\equnp{mcmig2}, ignoring accretion)} &
  $\sim 0.50$    &  0.73        &   0.54     &   0.68       &  -         &   $\sim 0.12$   &     -       \\
\hline
\end{tabular}
\end{minipage}
\label{tab:obs}
\end{table*}

\Tab{obs} lists pioneering estimates of  
the properties of seven giant clumps as observed in five 
$z \simeq 2.2$ star-forming disc galaxies (SFG) using AO spectroscopy focusing
on H$\alpha$ at the ESO VLT as part of the SINS survey \citep{forster09}.
These data are based on Table 2 of \citet{genzel11}, with slight revisions
for the massive clumps in ZC406690 from Table 3 of \citet{newman12}.
The five galaxies are selected to be massive discs of rotation velocities 
$\sim 250 \kms$,
dynamical masses of more than $10^{11}\msun$ within the inner $10\kpc$,
and SFR$\sim 120-290\sy$. They sample the upper end of the SFG population,
and therefore the most massive giant clumps.

\smallskip
The galaxies BX482 and ZC406690 are large clumpy rotating discs with a
prominent $\sim 5\kpc$ ring of clumps and star formation. 
D3a15504 is a large rotating disc with a central AGN.
ZC782941 is a more compact rotating disc, showing an asymmetry due to a
compact clump off the main body of the galaxy, potentially a minor merger.
BX599 is a compact system with a high velocity dispersion and a small
$\sim 3\kpc$ rotating disc.

\smallskip
The most prominent clumps were identified from at least two different maps of
H$\alpha$ velocity channels.
Clump 1 in \tab{obs} is the dominant clump (A) in BX482, 
part of a $\sim 5\kpc$ ring that includes 3 additional smaller clumps.
Clump 2 is an average over the 6 off-centre clumps (A-F) in D3a15504, 
none of which is particularly dominant over the others.
Clump 3, from ZC782941, is at the largest distance from the disc centre and 
the brightest in H$_\alpha$, while this galaxy shows four additional clumps 
closer to the centre.
Galaxy ZC406690 shows 4 clumps in a $\sim 5\kpc$ ring, of which three were 
studied in \citet{genzel11} and listed here.
Clump 4 is ZC406690-C, the brightest in I-band ACS, faint in H$_\alpha$,
and shows an elongated shape.
Clump 5 is ZC406690-A, the brightest in H$_\alpha$ and rather round, compact 
and isolated. Its SFR is high, its stellar population is young, and it is gas 
rich.
Clump 6 is ZC406690-B, the second in H$_\alpha$ brightness and 
rather faint in I-band ACS. Its stellar population is rather old, and it is 
relatively gas poor.
Clump 7 is an exception, the whole centre of the compact galaxy BX599, 
namely a compact star-forming bulge.

\smallskip
The first group of rows in \tab{obs} refer to the clump structural properties.
The second group of rows are the observed SFR and wind properties.
The third group is quantities deduced from the observed quantities.
The quantities marked by ``*" are directly deduced from the observations.

\smallskip
As described in \citet{genzel11} and \citet{newman12}, 
the quoted quantities are highly uncertain. 
They are limited by resolution and by modeling assumptions.
For example, the formal errors quoted in Table 3 of \citet{newman12} 
are about 100\% for some of the quantities characterizing the winds.
These pioneering observations should therefore serve as preliminary 
indications only.

\smallskip
The intrinsic clump radius $\Rc$ was determined from the HWHM of a Gaussian 
fit to the appropriate velocity channel after subtracting in quadrature 
the HWHM of the instrumental resolution. Since the latter is typically 
$2 \kpc$, larger than the intrinsic clump radius, the estimated $\Rc$ 
is rather uncertain. 
 
\smallskip
The clump characteristic velocity $\Vc$ is derived here from 
the kinematic measurements of velocity dispersion $\sigma$ and rotation $\Vrot$
assuming Jeans equilibrium:
$\Vc^2 = \beta (\Vrot^2 + c\,\sigma^2)$.
The steep clump density profiles dictate $c \simeq 3.4$ (R.~Genzel, private
communication), and $\beta \simeq 1.17$ \citep{genzel11}.
Then the dynamical clump mass is derived from $\Mc = G^{-1} \Vc^2 \Rc$.
This gives larger masses than derived in \citet{genzel11} using $c=2$,
the value appropriate for an isotropic isothermal sphere.
\citet{genzel11} evaluated the clumps' gas mass from the measured SFR
using an adopted version of the Kennicutt-Schmidt (KS) law. 
With the recent calibration at $z \sim 2$ using  
CO measurements \citep{tacconi13}, the Kennicutt relation is not very different
than \equ{sfr} with $\epsf \simeq 0.01$, and the estimated gas mass using the
KS law with the recent calibration is similar to the dynamical mass as 
derived here.
Note, however that the dynamical mass could be underestimated if
the clump deviates from equilibrium due to strong inflows or outflows. 
 
\smallskip
The SFR is derived from the H$_\alpha$ luminosity corrected for extinction, 
with an uncertainty of about 30\%.
The clump outflow velocity $\Vw$ is estimated from the maximum blue shift 
%\adb{
and width
%}
of the broad emission component -- this is the main pioneering discovery of
\citet{genzel11}. Its error is about 33\%. 
%\adb{
However, by adopting the maximum wind velocity as the characteristic
wind velocity $\Vw$ one may overestimate some of the calculated wind 
properties.
%}
The mass outflow rate $\Mdotw$ is taken as the average of two different crude 
estimates using two photodissociation case-B models, 
described in appendix B of \citet{genzel11}. 
Because of the elaborate modeling involved, and the different results obtained
from the different models, this quantity is naturally 
highly uncertain, with an error on the order of 100\%.
This is what makes the current results indicative only, not to be taken too
strictly on a case by case basis.

%--------------------------
\subsection{Comparison of theory to observations}

The first four clumps, addressed as ``typical" clumps,
seem to be consistent with the fiducial case discussed above for stellar
momentum-driven outflows.
The SFR efficiency $\epsf_{,-2}$ is of order unity,
the wind velocity is 2-4 times the clump velocity, $\nu \sim 3$,
the mass loading factor $\eta$ is about unity and in one case $\sim 2$,
and the momentum injection-ejection factor $\fw$ is 2-3 and in one case $\sim
4$, as predicted by the theoretical momentum budget discussed above.

\smallskip
The 5th clump, ZC406690-A, is unusual in terms of its high SFR of $40 \sy$, 
but is still marginally consistent with the fiducial momentum-driven wind case.
It shows a marginally high SFR efficiency of $\epsf_{,-2} \simeq 3.8$.  
Its outflow is on the strong side, with $\eta \simeq 2.9$ and $\fw \simeq 6$, 
but this is still marginally consistent with the fiducial case.
However, it is different in the sense that its high SFR and low $\tff$ 
yield a short depletion time of $\sim 60\Myr$, which is about half the 
migration time. This clump will complete its migration intact only because
the mass gain by accretion from the disc is expected to be larger than the
mass loss by outflow. 

\smallskip
The last two clumps seem to be extreme cases of strong outflows that are 
inconsistent with stellar feedback, even when the the adiabatic supernova
feedback is at its maximum and the ejection into the wind is efficient.
Clump 6 is the most extreme case in terms of outflow. 
It is the most massive clump, $\sim 10^{10}\msun$,
its SFR efficiency is rather typical, $\sim 11 \sy$, 
with the stellar population rather old,
but its outflow is excessive, with $\eta \simeq 7$ and $\fw \simeq 34$.
Based on our estimates of the momentum budget, such an outflow cannot 
be driven by stellar feedback. 
Either it requires another driving mechanism,
the observational estimates are severe overestimates,
or the SFR we measure today is substantially
smaller than it was when the outflow was launched.

\smallskip
The bulge clump 7 is an exception, representing a whole galaxy
rather than a Toomre clump embedded in a disc. 
It has the highest SFR$\simeq 66\sy$.
It shows an outflow with a moderately large $\eta \simeq 2.8$  
but with a very high momentum injection efficiency of $\fw \simeq 14$.
As a result, 
its depletion time of $\tdep \simeq 63\Myr$ is only a quarter of its migration
time.
According to our momentum budget, such a high value of $\fw$ is more than 
what stellar feedback can offer in clumps; again, one could avoid this problem 
if the SFR we measure today
is lower than it was when the bulk of the outflowing material was launched. 

\smallskip
\citet{andrews10} proposed a scenario where multiple scattering is possible
when the gas surface density is above a threshold of 
$\Sigma_{\rm gas} Z \sim 1.1 \gcms$.
If this was true, and if the gas surface density in the extreme clumps
was sufficiently high, this could have provided a possible explanation for the
extreme clumps.  
However, the simulations of \citet{krumholz13} show that multiple scattering
does not occur even at a very high surface density. 
Furthermore, 
we note that the gas surface density in all the observed clumps is in the range
$(0.2-0.8)\,\gcms$, below the suggested threshold value
given that the metallicity is comparable to and slightly lower than solar.
There seems to be a marginal correlation between $\Sigma_{\rm gas}$ and $\fw$,
but it does not help us explain the extreme outflows in clumps 6 and 7.

%%%%%%%%%%%%%%%%%%%%%%%%%%%%%%%%%
\section{Conclusion}
\label{sec:conc}

We have analyzed the outflows expected from star-forming giant clumps 
in high-$z$ disc galaxies that undergo violent disc instability (VDI).
We evaluated the outflow properties based on the momentum budget, namely
the efficiency of momentum injection into the ISM per unit star formation rate
by a variety of stellar momentum and energy sources.
We then estimated the lifetime of the clumps given their VDI-driven
migration toward the disc centre and the associated growth of clump mass 
by accretion from the disc.

\smallskip
Our results can be summarized as follows:

\bul
Most of the mass loss is expected to occur through a steady wind
over many tens of free-fall times, or several hundred Myr,
rather than by an explosive disruption in one or a few free-fall times,
less than $\sim 100 \Myr$.

\bul
Radiation-hydrodynamics simulations by \citet{krumholz12c} and 
\citet{krumholz13} provide a key input to the momentum budget,
that radiation trapping is negligible because it destabilizes the wind.
This means that each photon can contribute to the wind 
momentum only once, and the radiative force is limited to about $L/c$.
This calls into question other recent works that assume a very large 
trapping factor without self-consistently computing it.

\bul
All the direct sources of momentum taken together inject momentum into
the ISM at a rate of about $2.5\,L/c$.
This includes radiation pressure, protostellar winds, main-sequence winds and
direct injection of momentum from supernovae.

\bul
The early adiabatic phases in expanding supernova-driven shells 
and main-sequence winds, if they operate at maximum efficiency, 
bring it up to a total force of $5\,L/c$ 
for typical gas densities in the clumps.
An unknown fraction of this force is actually used to drive the wind,
so this can serve as an upper limit.

\bul
The resulting outflow mass-loading factor is of order unity.
If the clumps were allowed to deplete their gas into stars and outflows
standing alone,
the depletion timescale would have been a few disc orbital times, 
a significant fraction of a Gyr, ending with about half the original 
clump mass in stars.

\bul
However, the clump migration time to the disc centre due to the VDI 
is on the order of an orbital time, about $250 \Myr$, so the typical
clumps are expected to complete their migration prior to depletion.

\bul
%\adb{
Furthermore, based on analytic estimates and simulations,
the clumps are expected to double their mass in a disc orbital
time by accretion from the disc and mergers with other clumps,
which overwhelm the mass loss by tidal stripping.
This high rate of gravitational mass growth implies a net growth of
clump mass in time and with decreasing radius despite the continuous 
massive outflows.
%}

\bul
From the six disc clumps observed so far, five are consistent with the
predictions for stellar-driven outflows.

\bul
One extreme case shows an outflow with an estimated
mass-loading factor of 7 and a momentum injection rate of $34\,L/c$.
This may indicate that the observed outflow in this case is an overestimate,
which is not unlikely given the large uncertainties in the observed
 properties.  
Otherwise, this may hint to a stronger driving mechanism.
One possible way to obtain higher efficiencies is if the supernovae
explode in extremely low density environments generated by the other 
feedback mechanism. Another possibility is that this clump is
just now ending its star formation, and therefore the present
measured SFR is smaller than the value that prevailed at the time
most of the outflow was launched.

\smallskip
We conclude that stellar feedback is expected to produce steady  
massive outflows from the high-z giant clumps, with mass-loading factors of 
order unity and momentum injection rate efficiencies of a few.
This is consistent with 5 of the 6-7 observed giant clumps where outflows
were observed so far, with one or two exceptions in which the estimated 
outflows are apparently stronger.
Despite the intense outflows, which indicate gas depletion times of 
several hundred Myr, the clumps are not expected to disrupt by this process.
Instead, they are expected to migrate to the disc centre on a somewhat shorter 
timescale, roughly a disc orbital time or about $250 \Myr$
\citep{dsc09,cdb10}, 
and during this process they are expected to more than double in mass 
by accretion from the disc. 
One therefore expects the population of in-situ clumps
to show systematic variations in their properties as a function of radius in 
the disc, in the form of declining mass, stellar age and metallicity,
and increasing gas fraction and specific star-formation rate
\citep{ceverino12,mandelker13}.
One also expects that the clump migration, combined with the VDI-driven
inter-clump gas in the discs, is an efficient mechanism for forming compact
spheroids \citep{dsc09,cdb10,dekel13}, 
and providing fuel for black-hole growth and AGN activity \citep{bournaud11}.  

%%%%%%%%%%%%%
\section*{Acknowledgments}

We acknowledge stimulating discussions with Reinhard Genzel and the SINS team,
and thank the anonymous referee for a helpful report.
AD acknowledges support by ISF grant 24/12,
by GIF grant G-1052-104.7/2009,
by a DIP-DFG grant,
and by NSF grant AST-1010033.
MRK acknowledges support from the Alfred P.~Sloan
Foundation, from the NSF through CAREER grant
AST-0955300, and by NASA through a Chandra Space
Telescope Grant and through ATFP grant NNX13AB84G.

%%%%%%%%%%%%%%%%%%%%%%%%%%%%%%%%%%%%%%%%%%%%%%
%\bibliographystyle{mn2e}
%\bibliography{flows}

%%%%%%%%%%%%%%%%%%%%%%%%%%%%%%%%%%%%%%%%%%%%%%

%%%%%%%%%%%%%%%%%%%%%%%%%%%%%%%%%%%%%%%%%%%%%%55
\label{lastpage}
\end{document}